\documentclass{shep3}

\usepackage{amsmath,amssymb}
\usepackage{graphicx}
\usepackage{algorithmic,algorithm}
\usepackage{url}

\newcommand{\reduze}{{\tt Reduze}\;}

\preprint{ZU-TH 18/09}

\title{Reduze -- Feynman Integral Reduction in C++}

\author{C.~Studerus\footnote{Email: {\tt cedric@physik.uzh.ch}} \\

\noindent {\it Institut f{\"u}r Theoretische Physik,
Universit{\"a}t Z\"urich, Winterthurerstrasse 190,
CH-8057 Zurich, Switzerland}
}

\abstract{\reduze is a computer program for reducing Feynman Integrals to master integrals
employing a Laporta algorithm.
The program is written in C++ and uses classes provided by the {\tt GiNaC} library to
perform the simplifications of the algebraic prefactors in the system of equations.
\reduze offers the possibility to run reductions in parallel.}


\begin{document}
\hyphenation{Stu-de-rus}
%
%
%
\section{Introduction}
The calculation of loop amplitudes in perturbative quantum field theory is usually
done by generating the Feynman diagrams for the desired physical process and
interfering the corresponding analytical expressions, working out the Dirac
and/or color structure. The amplitude is then a sum of many dimensionally
regularized integrals~\cite{'tHooft:1972fi} which have to be computed.
These integrals are not independent of each other but related by the Integration
by Parts (IBP) Identities~\cite{Tkachov:1981wb,Chetyrkin:1981qh} and the Lorentz
Invariance (LI) Identities~\cite{Gehrmann:1999as}.
These identities form a homogeneous system of linear equations with the integrals
as unknowns and algebraic prefactors which are rational polynomials in the kinematic
invariants and the dimension.
Using the IBP and LI identities one can express most of the integrals in terms of
a small set of integrals, called master integrals.

These identities also exist for phase-space integrals after replacing the delta
functions by a difference of propagators with an opposite sign prescription
of the imaginary part~\cite{Anastasiou:2002yz}.

The procedure of solving this system of equations is called a reduction. Since one
often has to solve systems with thousands of equations, computers have to be used
and because the prefactors in front of each integral are rational polynomials,
computer algebra systems become indispensable.

\reduze is a computer program written in C++ which generates the IBP and
optionally the LI identities and then reduces the integrals to master integrals.
\reduze uses the {\tt GiNaC} library~\cite{Bauer:2000cp} to perform the simplification
of the prefactors.

The reduction algorithm is a Laporta algorithm~\cite{laporta} which is essentially
the Gauss algorithm with additional rules to determine the next equation which should be
solved and inserted into the others. To get the reduction of a certain Feynman diagram
one first defines a set of integrals by restricting the exponents of the propagators.
\reduze then generates the identities from this set and starts solving the
system of equations.

For a reduction of several diagrams \reduze can treat different diagrams (with the
same number of propagators) simultaneously. One defines how many cores or processors
are available and then \reduze will automatically launch some reductions simultaneously.
The more cores there are available the more diagrams one can reduce in parallel.

Other published reduction programs are AIR~\cite{Anastasiou:2004vj} and FIRE~\cite{Smirnov:2008iw}.
AIR is a {\tt Maple} package that implements the Laporta algorithm. FIRE is a
{\tt Mathematica} package that implements the Laporta algorithm and also a method that
uses techniques from Gr\"obner basis calculations. Then there are also several other
private codes.

The advantage of \reduze is that it is completely open source, has a low memory footprint
and can do reductions in parallel.
\subsection{License}
The package \reduze is Copyright \textcopyright \;2009 Cedric Studerus.
This program is free software: you can redistribute it and/or modify
it under the terms of the GNU General Public License as published by
the Free Software Foundation, either version 3 of the License, or
(at your option) any later version.

This program is distributed in the hope that it will be useful,
but WITHOUT ANY WARRANTY; without even the implied warranty of
MERCHANTABILITY or FITNESS FOR A PARTICULAR PURPOSE.  See the
GNU General Public License for more details.

You should have received a copy of the GNU General Public License
along with this program.  If not, see {\tt <http://www.gnu.org/licenses/>}.
%
%
%
\section{Theoretical background}
\subsection{Propagators, Sectors and Integrals}
A propagator $P$ with momentum flow $q$ and mass $m$ is the expression
$q^2 - m^2$ where $q^2$ denotes the Minkowski scalar product with the
metric in the convention $g=$ $\mbox{diag}$\! $(1,-1,-1,-1)$.
The momentum $q$ is a linear combination of loop momenta
$k_i$ and external momenta $p_i$.

An auxiliary topology is an ordered set of propagators $A_n = \{ P_1,\ldots, P_n \}$
such that all scalar products $k_i \, k_j$ and $k_i \, p_j$ containing at least one
loop momentum $k_i$ can be expressed by a linear combination of propagators from $A_n$.
The auxiliary topology is called an $l$-loop auxiliary topology if there are $l$ different
loop momenta appearing in the momenta $q$ of the propagators. Denoting the number of
independent external momenta by $m$, the auxiliary topology must contain exactly
$l\,(l+1)/2+l\,m$ propagators where the first term counts the scalar products
between loop momenta only and the second term the products involving both loop and
external momenta.

Every subset of $t$ propagators of a given auxiliary topology $A_n$ defines a
sector $T_t$
with an unique identification number $ID$. Physically relevant sectors which correspond
to diagrams are also called topologies. Assuming the sector has the
propagators $P_{j_1},\ldots, P_{j_t}$ with $\{j_1,\ldots,j_t\} \subset \{1,\ldots,n\}$,
then its identification number is defined as
\begin{equation}
 ID = \sum_{k=1}^{t} 2^{j_k-1} \, .
\end{equation}
There are in general $\binom{n}{t}$ different $t$-propagator sectors $T_t$ and
$\sum_{t=1}^{n} \binom{n}{t} = 2^n - 1$ sectors that one can build out of an
auxiliary topology $A_n$. Their identification numbers fulfill $1 \leq ID \leq 2^n - 1$ .

A sub-sector $T_{t-1}$ of a sector $T_{t}$ is a sector where one propagator is removed.
There are in general $t$ different sub-sectors for a
sector $T_{t}$. The sub-sector tree of a sector $T_t$ is the set of all sub-sectors
of $T_t$ and recursively all sub-sectors of all these sub-sectors.
All sectors of an auxiliary topology $A_n$, which is the main sector, are in the
sub-sector tree of $A_n$. \\

To every $t$-propagator sector $T_t$ with propagators $P_{j_1}$, $\ldots$, $P_{j_t}$
belongs a infinite set of $d$-dimensionally regularized $l$-loop
integrals~\cite{'tHooft:1972fi} which all share the same propagators in the denominator
of the integrand. They have the generic form
\begin{equation}\label{eq:dim_reg_int}
\int \mbox{d}^d k_1 \ldots \int \mbox{d}^d k_l \,
     \frac{P_{j_{t+1}}^{s_1} \ldots P_{j_n}^{s_{n-t}}}{P_{j_1}^{r_1} \ldots P_{j_t}^{r_t}}
\end{equation}
with integer exponents $r_i \geq 1$ and $s_i \geq 0$.
In \reduze such an integral is represented by a vector $v = \{v_1, \ldots, v_n\}$ containing
the exponents of the propagators or, more precisely,
\begin{equation}\label{eq:int_reduze_format}
\mbox{INT}[t, r, s, ID,\{v_1,\ldots,v_{n}\}]
\end{equation}
where $r=\sum_{i=1}^t r_i \geq t$ is the sum of the propagators in the denominator and
$s=\sum_{i=1}^{n-t} s_i \geq 0$ is the sum of the propagators in the numerator.
The value $v_i$ is the exponent of propagator $P_i \in A_n$. It is positive if $P_i$
is in the denominator, negative if $P_i$ comes with a (positive) exponent in the numerator
and zero if the propagator is not present. The numbers $t$, $r$, $s$ as well as the
identification number $ID$ of the sector, to which the integral belongs, can be
calculated from the vector $v$.

Consider an $n$-propagator auxiliary topology $A_n$ with a $t$-propagator sector $T_t$.
The number of integrals that one can build for certain values of $r$ and $s$ is given by
\begin{equation} \label{eq:nr_of_ints}
{\mathcal N}(n,t,r,s) = \binom{r-1}{t-1} \binom{s+n-t-1}{n-t-1} \, .
\end{equation}
The two binomial factors count all possible ways to arrange the exponents of
the propagators in the denominator and numerator, respectively.
\subsection{Integration By Parts (IBP) Identities}
In dimensional regularization~\cite{'tHooft:1972fi} the integral over a total derivative
is zero. Let ${\mathbf I'}$ be the integrand of an integral of the form (\ref{eq:dim_reg_int}).
Then, working out the differentiation in
%
\begin{equation}
 \int \mbox{d}^d k_{i} \, \frac{\partial}{\partial k_{i}^{\mu}} \, \big[ q^{\mu} \,
 {\mathbf I'}(p_1, \ldots, p_{m}, k_1, \ldots, k_l) \big] = 0
\end{equation}
leads to the integration by parts (IBP) identities~\cite{Tkachov:1981wb,Chetyrkin:1981qh}.
The momentum $q$ is an arbitrary loop or external momentum. The index $\mu$ is summed over
but the index $i$ is not. If there are $l$ loop momenta and $m$ independent external
momenta one can therefore build $l\, (l+m)$ equations from one integral (the seed integral).
\subsection{Lorentz Invariance (LI) Identities}
One can also use the Lorentz Invariance of the integrals~\cite{Gehrmann:1999as}. Taking
an integral ${\mathbf I}(p_{1},\ldots ,p_{m})$ the following equation holds
\begin{equation}
\sum_{n=1}^m \Big( p_{n}^{\nu} \frac{\partial}{\partial p_{n \mu}}
    - p_{n}^{\mu} \frac{\partial}{\partial p_{n \nu}} \Big)\, {\mathbf I}
    (p_{1},\ldots ,p_{m}) = 0 \, .
\end{equation}
The derivatives can be shifted directly to the integrand of the integral ${\mathbf I}$.
This equation can be contracted with all possible antisymmetric combinations of the
external momenta, e.g. $p_{1 \mu} p_{2 \nu} - p_{1 \nu} p_{2 \mu}$, which leads to
$m\,(m-1)/2$ equations where $m$ denotes the number of independent external momenta.
As it was shown in~\cite{Lee:2008tj} the LIs do not give new linear independent
equations in addition to the IBPs. However, they can accelerate the convergence in a
reduction, since in general an LI identity generated from one seed integral cannot be
reproduced with the IBP identities generated from the same seed integral alone.
\reduze offers the possibility to use the LIs.
\subsection{Symmetry Relations} \label{sec:sym_rel}
Often there are relations between integrals coming from symmetries which can
lead to an identification of integrals and even whole sectors.

All integrals are invariant under permutations of the loop momenta and translations
of a loop momentum with other momenta. Such a transformation can be used to transform
an integral $I$ to an equivalent integral $I'$ and leads to the identity $I=I'$.
However, since the integrals are expressed with propagators of an auxiliary topology,
the transformations that actually can be used must leave the set of propagators of
the auxiliary topology invariant. This means that such an symmetry transformation must
lead to a permutation of the propagators and then two equivalent integrals differ only
in a permutation of the propagator exponents.

A permutation of the propagator exponents of an integral can alter the sector (sector
identification number) it belongs to. This is then valid for all integrals of this sector
and one can completely get rid of one of the sectors. These sectors correspond to the
same topology.

Sometimes the integrals of an auxiliary topology are invariant under the permutation
of external momenta but this permutation leads to propagators which are not contained
in the auxiliary topology. One then has to find a transformation on the propagators
which transforms the propagators back into the auxiliary topology.

In \reduze one can explicitly declare some transformation rules which leave the
integrals invariant. Since only transformations that lead to a permutation of
propagators can be used, they must be given as permutations. According to these
declarations \reduze automatically identifies equivalent integrals in the system of
equations and also only uses one of them to generate the equations.
\subsection{Zero Sectors}
It is possible that a whole sector is zero which means that all integrals belonging
to this sector are zero. In \reduze a sector is set to zero if the solutions of
all IBP identities generated from the integral $I$ of this sector with $r=t$ and
$s=0$ (no additional propagators in the numerator and denominator) contain the equation
$I=0$.
\subsection{Reduction}
To reduce a sector up to $r=r_{max}$ and $s=s_{max}$ means solving the
homogeneous system of linear equations which have been built with the IBP
and/or LI identities out of the integrals of this sector that have
$r \in [r_{min}, r_{max}]$ and $s \in [s_{min}, s_{max}]$. It is not possible
to solve the system completely since the rank of every finite set of equations
is smaller than the number of unknown integrals it contains. However, the aim of
the reduction is to express most of the integrals by a linear combination of only
a few (less complicated) integrals, the so-called master integrals.

In a typical reduction of a $t$-propagator sector one takes the smallest
possible values $r_{min} = t$ and $s_{min} = 0$. If one then generates the equations
from all integrals with $r \in [r_{min}, r_{max}]$ and $s \in [s_{min}, s_{max}]$
and reduces this system of equations, one usually gets the solutions for all the
integrals that have been chosen to build the system, meaning that all these integrals
are expressed as a linear combination of some master integrals.

The equations built to reduce a sector can contain not only integrals of the sector
itself but also integrals from sub-sectors. If one tries to reduce the system of
equations, the results then still depend on a lot of unsolved integrals of the
sub-sectors which also have to be reduced by solving a system of equations built
with integrals from the sub-sector considered. Sub-sectors appear as long as the
number of propagators does not go below a minimal value.
\section{Reduction Algorithm}
For a reduction of a certain sector including its sub-sectors \reduze first
determines the whole sub-sector tree, then reduces the sub-sectors with the smallest
number of propagators, inserts the results in the equations of the sectors that
depend on these sub-sectors, reduces these sectors, continuing until the desired
sector with the largest number of propagators can be reduced. For a reduction of a
single sector \reduze first will generate the equations and insert the results of
the sub-sectors into them if they are available in the default {\tt results}-directory.

\reduze is not able to find a solution for a single integral of a sector, instead it
employs all integrals in a user-defined range of $r$ and $s$ for building the equations
and then reduces the whole system.

Since the reduction of a single sector often involves a huge amount of equations,
the system must be divided into smaller parts. Dividing the equations in subsets,
each generated from a set of integrals for a certain value for $r$ and $s$, can
still lead to systems which are too big.
In \reduze the system of equations is divided into smaller sets with a default
number of equations. The number of equations per set can be adjusted by the user.
To decide which equation goes in which set, the equations are first sorted in
descending order with respect to the most complicated integral each equation
contains and then simply divided into smaller sets and stored in temporary files.

The reduction of a sector is now done by first loading the temporary file
containing the simplest equations. Then this system is reduced according to the
algorithm below and the results are inserted in all the other files. Then
the equations from the second file will be reduced and the results again inserted
in all other files. This procedure is done for all files. Finally, the results
from the temporary files are collected and saved in a single result file.

To give a precise meaning for the instance that an integral is simpler or less
complicated than another integral a lexicographic ordering can be defined~\cite{laporta}:
For an integral
\begin{equation}
 I = \mbox{INT}[t, r, s, ID, \{v_1,\ldots, v_n\}]
\end{equation}
take the vector $\tilde{v}_I = \{t,r,s,ID, v_1,\ldots, v_n \}$ with length
$n+4$.
Then for two integrals $I$ and $J$ the comparison $I < J$ is true if and
only if there exists an $m \in \{1,\ldots,n+4\}$ such that
$\tilde{v}_I[m] < \tilde{v}_J[m]$ and $\tilde{v}_I[k] = \tilde{v}_J[k]$
for all $k \in \{1,\ldots, m-1\}$.

This operator can be extended to equations: Equation $e_1$ is less complicated than
equation $e_2$, $e_1 < e_2$, if the most complicated integral of $e_1$ is less
complicated than the most complicated integral of $e_2$.
For a set $L$ of equations that contains the integral $I$ in at least one equation
the set $[I]_L$ is defined as the subset of $L$ such that all equations have
$I$ as the most complicated integral.
In addition, the set $(I)_L$ is defined as the subset of equations of $[I]_L$ with
the smallest number of integrals. This set is the subset of $L$ which contains the
shortest equations and all equations contain the most complicated integral $I$ as
their own most complicated integral.
For an equation $e$ the expression $*e$ denotes the same equation but
solved for its most complicated integral.

The first part of the reduction algorithm (see Algorithm~\ref{alg:red}) is to bring
the system in a triangular form and the second part of the algorithm is the back
substitution.
\begin{algorithm}
 \caption{Reduction}
 \label{alg:red}
\begin{algorithmic}
\STATE $L = \{e_1,\ldots,e_n\}$	// list of equations.
\STATE $I = \{I_1,\ldots,I_m\}$		// all integrals in $L$, sorted in descending order.
\STATE $S = \{\}$			// empty list.
\STATE \COMMENT{ Triangularization:}
\FOR{$i = 1$ to $m$}
  \STATE choose $e \in (I_i)_L$
  \STATE $L = L\setminus{e}$
      \STATE replace matching integrals in $L$ by the r.h.s of $*e$
  \STATE $S = \{*e,S\}$
\ENDFOR
\end{algorithmic}
\begin{algorithmic}
\STATE \COMMENT{ $S = \{S(1),\ldots,S(l)\}$ contains now $l$ equations sorted in
    ascending order.}
\STATE \COMMENT{ Back substitution:}
\FOR{$i = 1$ to $l$}
  \STATE $e = S(i)$
  \FOR{$j = i+1$ to $l$}
    \STATE replace matching integrals in $S(j)$ by r.h.s of $*e$
  \ENDFOR
\ENDFOR
\end{algorithmic}
\end{algorithm}
%
%
%
\section{Usage}
\subsection{Finding an Auxiliary Topology}
Before a reduction can be launched the user has to find an appropriate auxiliary
topology which contains the topologies from given Feynman diagrams as sub-sectors.
The auxiliary topology should cover as many diagrams as possible and also allows for
as many as possible symmetry relations in order to minimize the number of
sectors which finally have to be reduced (see section~\ref{sec:sym_rel}).
The diagrams under consideration need a certain maximal amount of propagators.
By building these propagators one has some freedom how to choose the momentum flow and,
if the number of propagators of the diagram is smaller than the number of propagators
needed for the auxiliary topology, one must introduce additional auxiliary propagators.
This freedom of choosing how the propagators exactly look like and which additional
propagators are introduced can be used to set up an auxiliary topology that has
as many as possible symmetries.
\subsection{Reduction}
Since \reduze reads and writes a lot to files, it should be run on the local hard disc.
For the following it is assumed that the current working directory is a local directory.
A reduction is now done in three steps: set up an auxiliary topology, prepare the reduction
and run the reduction.
\subsubsection{Set up an Auxiliary Topology}
First one has to define an auxiliary topology, where momenta, propagators, etc. are declared.
For this one creates a file with suffix {\tt .in}, e.g. {\tt topoA.in}, with all the inputs
needed. The syntax of this input file is described in section (\ref{sec:aux_top_input}). There
are also some input files in the {\tt example} directory of the package. Then one sets up the
auxiliary topology (topoA) with the command
\begin{verbatim}
   reduze --setup topoA.in
\end{verbatim}
This creates the directory {\tt topoA} and initializes the auxiliary topology. It derives
the rules to express scalar products by propagators, finds the sectors that are
equivalent due to symmetry relations and finds most of the zero sectors
by doing a small reduction ($r = t$, $s = 0$) of all sectors which have vanishing
sub-sectors only.

The new created directory {\tt topoA} contains the directories {\tt reduction} and
{\tt results} and a log file {\tt setup.log}.
The log file contains some information about the auxiliary topology, e.g at the end of
this file there is a list of all sectors that have not been found to be zero and can be
reduced. The {\tt reduction} directory will later be used for the reduction. It contains
the directory {\tt include} in which the information about the auxiliary topology is stored.
One should not edit it directly. If the setup file {\tt topoA.in} was modified, the setup has to be done again. The {\tt results} directory is used for the results after
a reduction has completed.
\subsubsection{Prepare a Reduction}
To prepare a reduction one needs a second input file, e.g. {\tt prepareA.in},
in that one writes which sectors ($ID$ numbers) should be reduced and which
class of integrals ($R2=r_{max}, S2=s_{max}$) one wants to use for building the
system of equations. Also the maximum number of processes that will run in parallel
can be defined here.
The allowed commands of this file are described in section (\ref{sec:red_input}).
There are also some example files in the {\tt example} directory. Typing
\begin{verbatim}
   reduze --prepare prepareA.in
          --auxtop /path/to/topoA
\end{verbatim}
should evaluate very quickly. It initializes the input data as well as the inputs of
the auxiliary topology {\tt topoA}, checks them for consistency and creates the script {\tt run.sh}.
Every time one changes this input file it must be reprocessed with {\tt reduze --prepare}.
The script {\tt run.sh} will then be overwritten. The option {\tt --auxtop} followed
by an absolute or relative path to the directory of the auxiliary topology which is going
to be used can be omitted if one already works in the directory of this auxiliary topology.

The input data for the reduction are copied to the directory {\tt reduction/include}.
Avoid editing them directly.
\subsubsection{Run the Reduction}
The last step is the reduction. To start it one launches
\begin{verbatim}
   ./run.sh
\end{verbatim}
This script starts the reduction of the sectors declared in the input file {\tt prepareA.in}
and controls the number of processes running in parallel.
For every sector that has to be reduced a new directory will be created in the
directory {\tt reduction} with the identification number of the sector as name.
This directory is used for log files and temporary data.
When a reduction is complete the results are written to the directory {\tt results}.
If there are already results for this sector they will be overwritten.

The script {\tt run.sh} copies the executable {\tt reduze} from the installation
directory and renames it to {\tt reduzeID}, where ID is the identification number of the
sector that has to be reduced.
If one kills this script with
\begin{verbatim}
   killall run.sh
\end{verbatim}
the reductions of the sectors that have been started already will continue, but no
further reductions are started anymore. Single reduction processes can be killed with
\begin{verbatim}
   killall reduzeID
\end{verbatim}
where ID corresponds to the sector number.
\subsection{Manipulating the Results}\label{section:manipulate}
\reduze writes the results of a reduction in an internal format which is not well suited for
further processing with another algebra system. Also, if symmetry relations are used, the
result-files only contain the solutions for one of all equivalent integrals.
To get all the desired solutions the user must provide a list of integrals for which the solutions
should be extracted or generated due to symmetry relations. Then these results can be converted to
a {\tt Mathematica}- or {\tt FORM}-readable format \cite{mathematica,FORM}.

To do this, \reduze offers the options {\tt --select}, {\tt --FORM} and {\tt --MMA}.
These options should be used together with the option {\tt --auxtop} to tell \reduze which auxiliary
topology is going to be used.
\subsubsection{Select solutions}
\reduze can pick the solutions for some user defined integrals from all the results
generated during the reduction. Launching
\begin{verbatim}
   reduze --select MyIntegrals
\end{verbatim}
where {\tt MyIntegrals} is a file containing a semicolon separated list of
the integrals for which one wants to extract the solution. The format of the integrals
is the same as in formula (\ref{eq:int_reduze_format}), however, the numbers $t$, $r$,
$s$ and $ID$ can be omitted or replaced by any other character, which does not match the
bracket '\{'. A minimalistic example looks like
\begin{verbatim}
    INT[{v1, ..., vn}];
\end{verbatim}
%
The output is written to the file {\tt MyIntegrals.sol}. Integrals for which no solution
has been found (e.g. master integrals) will be written to the file {\tt MyIntegrals.rest}.
\subsubsection{Convert to {\tt FORM} and {\tt Mathematica} format} \label{sec:convert_to_FORM_MMA}
Invoking {\tt reduze} with the options
\begin{verbatim}
   reduze --FORMAT <files>
\end{verbatim}
where {\tt --FORMAT} is either {\tt --FORM} or {\tt --MMA} converts results from \reduze
format to a {\tt FORM} or {\tt Mathematica} readable format respectively.
The parameters given in {\tt <files>} are interpreted as the names of files containing
results in the format of \reduze\!.
For each input file a new output file is created in the current working directory with the extension
{\tt .inc} and {\tt .m} respectively. Usually, one first uses {\tt reduze --select} to extract only
the results that one actually needs and then converts them to the desired output format.

The {\tt Mathematica} format is a list of rules and the {\tt FORM}-format is a table
of {\tt id} statements. Positive integer powers of non-integer denominators in the prefactors of the
{\tt FORM} output are written with the function {\tt Den}, where {\tt Den(a)\textasciicircum k}
equals to $a^{-k}$.

Often one wants to expand the results, or more precisely the prefactors, in a Laurent series around $d-4$.
With \reduze one can do this expansion on the fly when creating the {\tt Mathematica} and {\tt FORM} outputs.
If such an expansion is desired, the user has to declare this in the reduction input file. The syntax is
explained in section (\ref{sec:output_options}).
%
%
%
\section{Input Files}
\subsection{Auxiliary Topology Input File} \label{sec:aux_top_input}
The input file, which defines an auxiliary topology, is read in to the main program at run
time. It must have the suffix {\tt .in}.
The input file consists of lines starting with a keyword followed by some values and a
semicolon. Keywords and values are separated by white spaces or tabulators.
Comments that will not be interpreted by \reduze are entered after a double slash and
empty lines are ignored.
\begin{verbatim}
    Keyword value; // don't forget the ';'
\end{verbatim}

In the following the keywords are explained using an explicit example.
The first declarations define the symbols and must be written at the very beginning
of the input file.
\begin{verbatim}
    LoopMomenta k1 k2;
    ExtMomenta p1 p2 p3;
    Symbol s t m;
    Dimension d;
\end{verbatim}
The keyword {\tt LoopMomenta} defines the loop momenta $k_1$ and $k_2$, {\tt ExtMomenta} the
external momenta $p_1$, $p_2$, $p_3$ and {\tt Symbols} declares the symbols
$s$, $t$ and $m$. {\tt Dimension d} sets the name of the dimension $d$; however,
if this line is missing the default is also taken to be $d$.

The keyword {\tt Propagator} takes as the first parameter the momentum flow and as the second
parameter the mass (not squared). In \reduze a propagator with momentum $q$ and mass $m$ is
the term $q^2 - m^2$. If one wants to work with the convention of the metric
$g = \mbox{diag}(-1,1,1,1)$, and therefore wants to have a propagator $q^2 + m^2$
one has to replace the mass {\tt m} by {\tt I*m}.
\begin{verbatim}
    Propagator k1 0;
    Propagator k2 0;
    Propagator k1-k2 0;
    Propagator k1-p1 0;
    Propagator k2-p1 0;
    Propagator k1-p1-p2 0;
    Propagator k2-p1-p2 0;
    Propagator k1-p3 m;
    Propagator k2-p3 m;
\end{verbatim}
These lines define nine propagators of which two have the mass $m$. The order of
the declarations is the same order that will be used in the integral representation
$\mbox{INT}[t,$\! $r,$\! $s,$\! $ID,$\! $\{v_1, \ldots ,v_n\}]$, where $v_i$ is the
exponent of the $i$-th propagator declared above.
There must be nine propagators because there are nine scalar products containing
loop momenta: $k_i^2, k_1\, k_2, k_i\, p_j$ for $1 \leq i \leq 2$ and $1 \leq j \leq 3$.
If these scalar products cannot uniquely be replaced by propagators \reduze will abort.

For diagrams with cuts only the (sub-)sectors which still contain all cut propagators
in the denominator of the integrand are non-zero. In other words: for diagrams with cuts one
wants to set to zero all integrals, which either miss one or more cut propagators
completely or where a cut propagator appears in the numerator of the integrand only.
To achieve this one declares the cut propagators with the keyword {\tt CutPropagator}
instead of {\tt Propagator}.

With the keyword {\tt Kinematic} all Lorentz scalar products between the external momenta
must be replaced by algebraic expressions of the the symbols defined above, otherwise
\reduze will abort.
\begin{verbatim}
    Kinematic p1^2 = 0;
    Kinematic p2^2 = 0;
    Kinematic p3^2 = m^2;
    Kinematic p1*p2 = s/2;
    Kinematic p2*p3 = (s+t-m^2)/2;
    Kinematic p1*p3 = (m^2-t)/2;
\end{verbatim}
\reduze will also abort if the left hand side of the equations is not a product of
exactly two external momenta or not a squared external momentum or if the right hand
side still contains some loop or external momenta.
For the process of two incoming massless particles with momenta $p_1$ and $p_2$ to
two outgoing massive particles with momenta $p_3$ and $p_4 = p_1 + p_2 - p_3$ with
squared mass $p_3^2 = p_4^2 = m^2$ the scalar product between different momenta are
replaced by the Mandelstam invariants $s = (p_1+p_2)^2$ and $t = (p_1 - p_3)^2$ in
this example.

There are several additional options one can give. One can set a variable, e.g. the mass
$m$, equal to one.
\begin{verbatim}
    SetToOne m;
\end{verbatim}
to reduce the number of variables in the rational polynomials by one.

Some integrals with different propagators are equal because the integrands
only differ by a shift of a combination of loop/external momenta or a permutation of
loop momenta. Those transformations that leave the set of propagators of the auxiliary
sector invariant can be given explicitly. The notation is adapted from the cycle
representation of permutations of the $n$ propagators and starts with the keyword
{\tt Permutation} followed by a sequence of the keyword {\tt Cycle} or {\tt Cyc}
with values from $\{1,\ldots,n\}$.
\begin{verbatim}
    Permutation Cyc 1 6 Cyc 2 7;
    Permutation Cyc 1 2 Cyc 4 5 Cyc 6 7 Cyc 8 9;
\end{verbatim}
The first line declares that there is a symmetry transformation which transforms
propagator 1 into 6, 6 into 1, and 2 into 7 as well as 7 into 2. This
transformation is the translation $k_i \to -k_i + p_1 + p_2$, $i=1,2$,
together with the exchange of the external momenta $p_1$ with $p_2$ and
$p_3$ with $p_4 = p_1 + p_2 -p_3$. The exchange of the external momenta
is allowed since it does not alter the Mandelstam invariants $s$ and $t$
on which the integral actually depends.
The second line implements the invariance of interchanging $k_1$ with $k_2$.
\reduze automatically finds all combinations of the user defined permutations.
In the above case the combination of the two permutations
\begin{verbatim}
    Permutation Cyc 1 7 Cyc 2 6 Cyc 4 5 Cyc 8 9;
\end{verbatim}
will automatically be added.

The setup procedure automatically determines the zero sectors. Setting the
\begin{verbatim}
    SetupFindZeros true;
\end{verbatim}
on {\tt false} will turn this off.
User defined zero or non-zero sectors can be declared explicitly with the keywords
\begin{verbatim}
    SetupZeroTopos    ID1 ID2     IDn;
    SetupNonZeroTopos ID1 ID2     IDn;
\end{verbatim}
where the keywords are followed by the desired sector identification numbers.
\subsection{Reduction Input File} \label{sec:red_input}
To prepare a reduction one creates a file in the directory of the auxiliary
topology. It must have the suffix {\tt .in}.
The following minimal declarations are necessary to reduce single sectors,
where all integrals with $t = R1 \leq r \leq R2$ and
$0 = S1 \leq s \leq S2$ are used to generate the equations. This is usually
enough to find all solutions of this class of integrals, but the results
will still depend on integrals of sub-sectors if they have not been
reduced before already.
\begin{verbatim}
    ReduceID 387 385 384 182;
    R2 5;
    S2 1;
\end{verbatim}
The keyword {\tt ReduceID} can be followed by as many sector identification
numbers one wants to reduce, the ordering of the numbers plays no role.
\reduze always begins reducing the sectors with the smallest number $t$ of
propagators. It automatically identifies sub-sectors and looks in the directory {\tt results}
for solutions for the sub-sectors. If such solutions exist, they will be inserted in the
equations of the sector before the reduction starts.
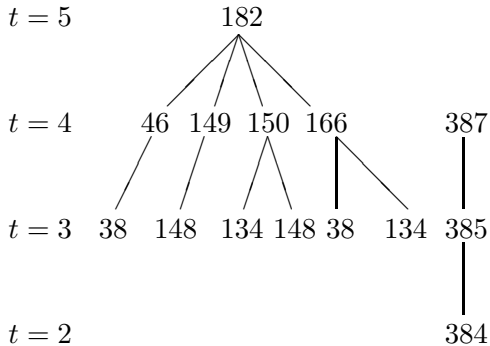
\begin{figure}[ht!]
\begin{center}
\begin{picture}(165,120)(0,8)
\put(80,120){182}
\put(87,117){\line(-1,-1){27}}
\put(87,117){\line(-1,-3){9}}
\put(87,117){\line(1,-3){9}}
\put(87,117){\line(1,-1){27}}
\put(50,80){46}
\put(68,80){149}
\put(90,80){150}
\put(112,80){166}
\put(54,78){\line(-1,-2){13.5}}
\put(74,78){\line(-1,-3){9}}
\put(98,78){\line(-1,-3){9}}
\put(98,78){\line(1,-3){9}}
\put(124,78){\line(0,-1){27}}
\put(124,78){\line(1,-1){27}}
\put(34,40){38}
\put(55,40){148}
\put(80,40){134}
\put(100,40){148}
\put(120,40){38}
\put(142,40){134}
\put(165,80){387}
\put(165,40){385}
\put(165,0){384}
\put(172,78){\line(0,-1){27}}
\put(172,38){\line(0,-1){27}}
\put(0,120){$t=5$}
\put(0,80){$t=4$}
\put(0,40){$t=3$}
\put(0,0){$t=2$}
\end{picture}
\end{center}
\caption{Sub-sector trees of sectors 182 and 387}
\label{fig:topotree182}
\end{figure}

If one wants to reduce some sectors together with their whole trees of sub-sectors
they depend on, one can use the keyword {\tt ReduceIDRecursive} followed by the
identification numbers. E.g. for the 5-propagator sector 182 and and the 4-propagator
sector 387, see Figure~(\ref{fig:topotree182}), one simply writes
\begin{verbatim}
    ReduceIDRecursive 182 387;
\end{verbatim}
This will first reduce the 2-propagator sector 384 (sub-sector of sector 385), then
the 3-propagator sectors 38, 134, 148 (from the sub-sector tree of sector 182) as well
as sector 385, then the 4-propagator sub-sectors of sector 182 and sector 387 and
finally sector 182. In this example the two sub-sector trees of 182 and 387 do not
overlap, but even if they do, each sector is only reduced once.

If one wants to exclude some sectors from the reduction, for example sector 166, which
is a sub-sector of sector 182, it can be done with
\begin{verbatim}
    ReduceIDNot 166;
\end{verbatim}
Note that the two sub-sectors (38 and 134) of sector 166 still will be reduced.

Excluding a whole sub-sector tree is done with
\begin{verbatim}
    ReduceIDRecursiveNot 166;
\end{verbatim}
This will exclude sector 166 and its whole sub-sector tree, in this case sector 38 and
134, from the reduction.

The commands {\tt ReduceID} and {\tt ReduceIDRecursive} for including some sectors
as well as {\tt ReduceIDNot} and {\tt ReduceIDRecursiveNot} for excluding
some sectors can have as many sectors as arguments as one wants.
One can also combine and repeat them as one likes, but one should notice that if a sector
is excluded with one of the exclusion commands it cannot be added again with another command.
Adding a sector multiple times has no more effect than adding it once.

There is also the possibility to set $R1$ and $S1$ to another value than $t$ and 0
respectively.
However, one cannot write $R1 = t$, since $t$ depends on the sector. Therefore the
definition $R1 = 0$ (the default) is interpreted as $R1 = t$ with the $t$ of the sector
under consideration, but any other declaration $R1 = n$ with $n > 0$ is taken literally.

The equations build from the integrals can be chosen with
\begin{verbatim}
    UseIBP true;
    UseLI  false;
\end{verbatim}
With this setting, the default, only the IBPs are used.

If the computer used for the reduction has several processors then one can declare
\begin{verbatim}
    Processes n;
\end{verbatim}
to tell \reduze the number of processes it can run simultaneously. Then, if a reduction is
launched with more than one sector with the same $t$, \reduze\! attempts to run $n$ sectors
in parallel.

The system of equations can become quite large, causing memory swapping.
To avoid this the system of equations is stored in several files, each holding a certain number of
equations, and then only the equations of a single file are loaded into memory and reduced.
With the following command one can change the number of the equations per file
\begin{verbatim}
    NrofEqperFile n;
\end{verbatim}
where $n=500$ is the default.
\subsubsection{Options for the {\tt FORM} and {\tt Mathematica} output}\label{sec:output_options}
%
%
%
In generating the {\tt FORM} and {\tt Mathematica} results one has
%
%
the possibility to expand the coefficients
in front of each integral in a Laurent series around $\epsilon = 0$, where $\epsilon$
is usually defined by $d=4-2\,\epsilon$. One has to tell \reduze the name of $\epsilon$
and the relation to the dimension $d$.
The following two commands are the default, if omitted.
\begin{verbatim}
    Epsilon ep;
    DimensionRule d = 4-2*ep;
\end{verbatim}
The next command tells \reduze actually to do this expansion.
\begin{verbatim}
    Series n;
\end{verbatim}
The integer $n$ is the order up to which the coefficients should be expanded,
including the order $n$ term. To generate the output see section (\ref{sec:convert_to_FORM_MMA}).
%
%
%
%
\section{Installation}
\subsection{Prerequisites}
\reduze uses the {\tt GiNaC} library~\cite{Bauer:2000cp} for the algebraic manipulations.
One must install {\tt GiNaC} version 1.4.1 or higher.
If the user's Linux distribution provides a compiled package of {\tt GiNaC}, one simply
can install the library and headers with the package manager.
Usually, if one wants to use a newer version of {\tt GiNaC}, it has to be compiled and
installed by hand.
\subsection{Building \reduze}
%
The most recent version of \reduze can be found at \url{http://www.itp.uzh.ch/~cedric/reduze/}.
Uncompress the package with
\begin{verbatim}
    tar -xzf reduze-version.tgz
\end{verbatim}
where {\tt version} is a placeholder for the current version of \reduze. Change to the
directory {\tt reduze-version} and configure, build, check and install \reduze with the
commands
\begin{verbatim}
    ./configure --prefix=/path/to/inst
    make
    make check
    make install
\end{verbatim}
The {\tt --prefix} option can be given if one wants to install \reduze in
{\tt /path/to/inst/bin} rather than in the default {\tt /usr/local/bin}.
The command {\tt make check} sets up an auxiliary topology, does a short reduction and
checks if these results are correct by comparing with an internal result file.

The installation directory must be appended to the {\tt PATH} environment variable.
If {\tt bash} is used, one writes in the profile file {\tt .bashrc}
\begin{verbatim}
    export PATH=/path/to/inst/bin:$PATH
\end{verbatim}
\reduze then can be invoked by typing
\begin{verbatim}
    reduze -h
\end{verbatim}
which gives a list of options.
%
%
%
\section{Performance}
Because of internal memory reasons the maximal number of propagators
is limited to $N=16$. The sum of the exponents of the propagators in
the numerator and denominator are limited to $R=16$ and $S=8$ respectively.

In a reduction most of the time is used for the algebraic manipulations on the
prefactors of the system of equations which are rational polynomials.
Any additional variable in these polynomials results in larger expressions
and makes the calculations slower. It is therefore very useful to put one
scale to one.

\reduze is implemented in a way that from a given set of integrals all IBP
identities are generated and then the whole system is reduced.
This system contains a lot of redundant equations, which lead to equations $0=0$
during the reduction. If the auxiliary topology allows for symmetry
transformations then one cannot only get rid of whole sectors but also some
integrals in a specific sector can be identified. This can drastically
reduce the number of equations.
%
%
%
\section{Applications}
The program \reduze was used to calculate the leading color coefficient
and the fermionic corrections to top-quark pair production in the quark-antiquark
channel at NNLO~\cite{studerus:fermion,studerus:planar}. For the leading color
coefficient one needs the reduction of the two planar box diagrams in
Figure~(\ref{fig:2loopboxes}). These are four-point functions depending on the
Mandelstam invariants $s,t$, the top-mass $m$ and the dimension $d$.
Including all the sub-sectors there are 60 sectors from which the reduction
identities of integrals up to three propagators in the numerator are needed.
These about 78'000 integrals are used to generate about half a million IBP identities.
The reduction of these equations then gives the solutions of all the 78'000 integrals
in terms of 35 master integrals. The running time for this reduction using 10
processors (2300 MHz) takes about 30 hours.
\FIGURE{
\centering
  \includegraphics[scale=0.4]{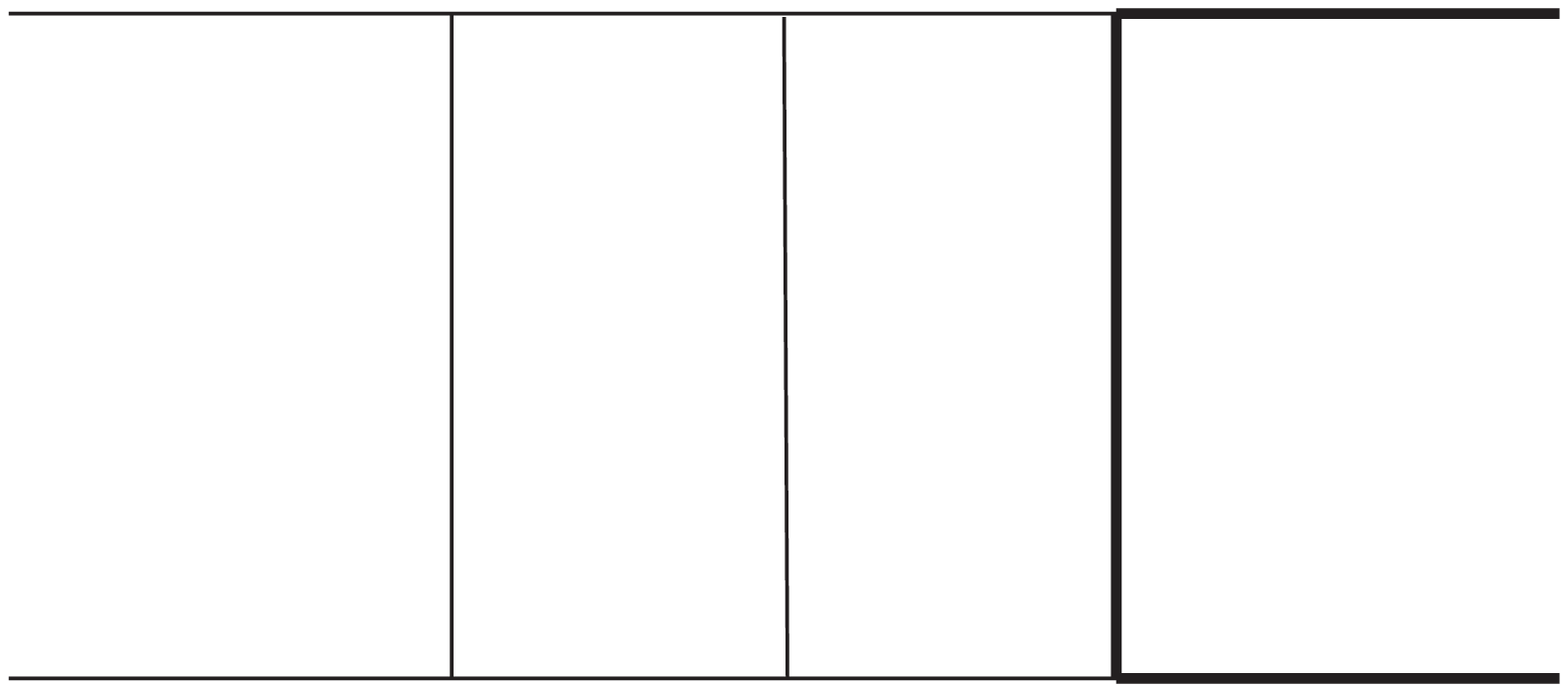}
  $\quad$
  \includegraphics[scale=0.4]{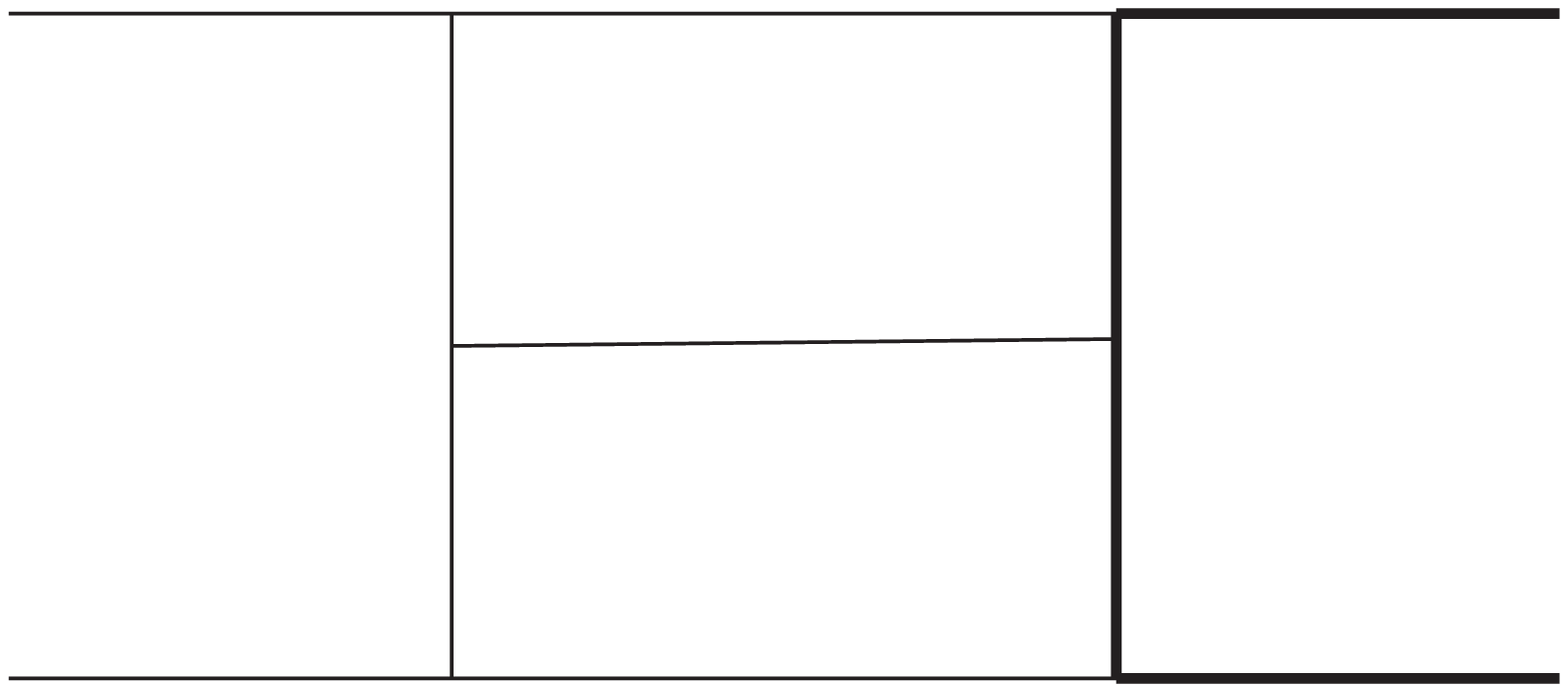}
 \caption{2-loop boxes for the process $q {\bar q} \to t {\bar t}$.}
 \label{fig:2loopboxes}
}
%
%
%
\section*{Acknowledgements}
The author wants to thank A. Ferroglia, T. Gehrmann, D. Ma\^{\i}tre, A. von Manteuffel
and D. Potter for many useful discussions. This research was supported by the
``Forschungskredit der Universit\"at Z\"urich''.
%
%
%
%

%
%
\end{document}